\documentclass[epj]{webofc}\woctitle{CONF12} % HEPHY-PUB 977/16
\usepackage[varg]{txfonts}\usepackage{amsmath}\usepackage{sidecap}
\begin{document}\title{On the description of exotic hadron states
within QCD sum rules}\author{Wolfgang Lucha\inst{1}\fnsep
\thanks{\email{Wolfgang.Lucha@oeaw.ac.at}}\and Dmitri Melikhov
\inst{1,2,3}\fnsep\thanks{\email{dmitri_melikhov@gmx.de}}}
\institute{Institute for High Energy Physics, Austrian Academy of
Sciences, Nikolsdorfergasse 18, A-1050 Vienna, Austria\and Faculty
of Physics, University of Vienna, Boltzmanngasse 5, A-1090 Vienna,
Austria\and D.~V.~Skobeltsyn Institute of Nuclear Physics,
M.~V.~Lomonosov Moscow State University, 119991, Moscow, Russia}

\abstract{Revisiting the QCD sum-rule description of exotic hadron
states, we argue that, in order to arrive at trustable analyses
of, for instance, strong decay widths of multiquarks, it is
inevitable to adopt the QCD sum-rule approach beyond the leading
perturbative level.}\maketitle

\section{Incentive: analytic nonperturbative approach to multiquark
bound states}Exotic hadrons are colour-singlet multiquark bound
states not composed just of a quark--antiquark pair or of three
(anti-) quarks. Candidates for this kind of multiquark bound
states, specifically, for mesonic four-quark states, so-called
tetraquarks, and for baryonic five-quark states, so-called
pentaquarks, have been observed by experiment. Here, we would like
to add a few general considerations on peculiarities and
opportunities encountered in an analysis of exotic hadrons by
means of QCD sum-rule techniques.

After recalling briefly the conceptual foundations and assumptions
of the QCD sum-rule approach, we point out a crucial qualitative
difference between the treatment of ordinary hadrons and the study
of exotic hadrons and we explore options for disclosing these
states' structure from their decay constants.

\section{QCD sum-rule formalism: the example of two-point
correlation functions}QCD sum rules \cite{SVZ,CK,BLI} are analytic
relationships between observable properties of a given hadron $H$
and the basic parameters of the quantum field theory of the strong
interactions, quantum chromodynamics. The starting point of their
derivation is a conveniently chosen operator which interpolates
the hadron $H$ under study, \emph{i.e.}, which has a non-vanishing
matrix element when being sandwiched between the QCD vacuum state,
$|\Omega\rangle,$ and the hadron's state $|H\rangle.$ Usually,
this interpolating operator is formulated as some current, defined
in terms of the fundamental degrees of freedom of QCD, quarks and
gluons, and, upon suppressing notationally the behaviour under
Lorentz transformations, henceforth generically~denoted by $j(x).$
The magnitude of its vacuum--hadron amplitude may be characterized
by a decay constant,~$f_H$:$$\langle\Omega|j(0)|H\rangle\propto
f_H\ne 0\ .$$The QCD sum rules then result from the evaluation of
appropriate $n$-point correlation functions of~such interpolating
currents in parallel at the level of hadrons and at the level of
the QCD degrees of freedom.\pagebreak

The clearly least intricate case is the \emph{two-point
correlation function\/}, given by the Fourier transform of the
(QCD-) vacuum expectation value of the time-ordered product of two
$H$-interpolating currents~$j,$ $$\Pi(p^2)\equiv{\rm i}\int{\rm
d}^4x\exp({\rm i}\,p\,x)\left\langle\Omega\left|{\rm
T}\!\left(j(x)\, j^\dag(0)\right)\right|\Omega\right\rangle\,.
$$Given the validity of the \emph{operator product expansion\/}
hypothesized by Wilson \cite{KGW,NSVZ}, a nonlocal product of
operators can be converted into a series of (linearly independent)
local operators ${\cal O}_n,$ $n=0,1,2,\dots,$ with Wilson
coefficient functions $C_n,$ depending on space-time distances,
$x^2,$ and an energy scale $\mu$ that defines the distinction of
perturbative and nonperturbative contributions to vacuum
expectation~values:$${\rm
T}\!\left(j(x)\,j^\dag(0)\right)=\sum\limits_{n=0}^\infty
C_n(x^2,\mu)\!:\!\!{\cal O}_n(x=0,\mu)\!\!:\:=C_0(x^2,\mu)+
\sum\limits_{n=1}^\infty C_n(x^2,\mu)\!:\!\!{\cal O}_n(x=0,\mu)
\!\!:\ .$$Herein, colons emphasize the free-field normal ordering
of the composite operators ${\cal O}_n(x,\mu),$ and in the second
equality the unit operator has been highlighted by separation from
the sum over local operators. The correlation function $\Pi(p^2)$
reflects the latter general expression of the operator product
expansion: $\Pi(p^2)$ consists of a portion $\Pi_{\rm
pert}(p^2,\mu)$ entirely determined by perturbative quantum
chromodynamics,$$\Pi(p^2)=\Pi_{\rm pert}(p^2,\mu)+
\sum_{n=2}^\infty\frac{C_n}{(p^2)^n}\,\langle\Omega|\!:\!\!{\cal
O}_n(0,\mu)\!\!:\!|\Omega\rangle\,,$$and a series of terms
controlled by the vacuum expectation values of all gauge-invariant
local operators ${\cal O}_n(x,\mu)$ that are constructible in
terms of the QCD fields and their derivatives, the \emph{vacuum
condensates\/}$$\langle\Omega|\!:\!\!{\cal O}(0,\mu)\!\!:\!
|\Omega\rangle\ne0\ ,$$which assume non-vanishing values as a
consequence of the nontrivial nature of the QCD vacuum,
$|\Omega\rangle.$ Allowing for complex values of the momentum
variable $p^2,$ taking advantage of the analytic behaviour of
$\Pi(p^2)$ as function of $p^2,$ performing in the complex-$p^2$
plane both the analytic continuation of $\Pi(p^2)$ and a contour
integration, exploiting the Cauchy integral formula, and assuming
an appropriately~rapid decrease of $\Pi(p^2)$ in the limit
$|p^2|\to\infty$ of the absolute value $|p^2|$ enables us to
represent the correlation function $\Pi(p^2)$ in the guise of a
\emph{dispersion relation\/} involving a spectral density
generically called $\rho(s)$:$$\Pi(p^2)=\int\frac{{\rm
d}s}{s-p^2}\,\rho(s)\ .$$At the hadron level, inserting a complete
set of intermediate hadron states into our vacuum expectation
value and singling out the hadronic state of interest,
\emph{i.e.}, the ground state of mass $M_H,$ by subsuming~all
excited and continuum states in a ``continuum'' contribution,
$\rho_{\rm cont}(s),$ yields the hadron spectral density
$$\rho_{\rm hadr}(s)=f_H^2\,\delta(s-M_H^2)+\rho_{\rm cont}(s)\
.$$At the QCD level, the spectral density inherits the partition
into perturbative and nonperturbative~parts:$$\rho_{\rm QCD}(s)=
\rho_{\rm pert}(s,\mu)+\sum_{n=2}^\infty C_n\,\delta^{(n)}(s)\,
\langle\Omega|\!:\!\!O_n(0,\mu)\!\!:\! |\Omega\rangle\ .$$At the
hadron level, removal of possible subtraction terms and reduction
of the impact both of hadronic excitations above the ground state
and of the hadronic continuum may be accomplished by applying, to
the correlation function, a \emph{Borel transformation\/} ${\cal
B}_\tau$ to a new variable, the Borel parameter $\tau,$
defined~by$$\Pi(p^2)\xrightarrow{\rm Borel}\Pi(\tau)={\cal
B}_\tau\!\left[\Pi(p^2)\right]\equiv
\lim_{\underset{\scriptstyle-p^2/n=1/\tau}{\overset{\scriptstyle
p^2\to-\infty}{\scriptstyle n\to\infty}}}\,\frac{(-p^2)^{n+1}}{n!}
\left(\frac{{\rm d}}{{\rm d}p^2}\right)^n\Pi(p^2)\ .$$\newpage
\noindent Of particular interest are the Borel transforms of poles
at a point $a,$ emerging from the above definition:$${\cal
B}_\tau\left[\frac{1}{(a-p^2)^k}\right]=\frac{\tau^{k-1}}{(k-1)!}
\exp(-a\,\tau)\ ,\qquad k\in\mathbb{N}_{>0}\equiv\{1,2,3,\dots\}\
.$$The ``continuum'' spectral density $\rho_{\rm cont}(s)$
vanishes below some physical threshold $s_{\rm phys},$ defined by
the location of the lowest-lying hadron excitation or the onset of
the hadronic continuum. The perturbative spectral density
$\rho_{\rm pert}(s,\mu)$ vanishes below some ``theoretical''
threshold $s_{\rm th}.$ For an interpolating current formed by a
quark and an antiquark, of masses $m_1$ and $m_2,$ respectively,
$s_{\rm th}$ is given by $s_{\rm th}=(m_1+m_2)^2.$ Therefore,
referring by $\Pi_{\rm NP}(\tau,\mu)$ to the Borel transform of
the total nonperturbative contribution to the QCD spectral density
$\rho_{\rm QCD}(s),$ our Borel-transformed (or ``Borelized'')
correlation function $\Pi(\tau)$~reads\begin{align*}\Pi(\tau)&=
\int{\rm d}s\exp(-s\,\tau)\,\rho(s)\\&=f_H^2\exp(-M_H^2\,\tau)+
\int\limits_{s_{\rm phys}}^\infty{\rm d}s\exp(-s\,\tau)\,\rho_{\rm
cont}(s)&&\mbox{(at hadron level)}\\&= \int\limits_{s_{\rm
th}}^\infty{\rm d}s\exp(-s\,\tau)\,\rho_{\rm pert}(s,\mu)+\Pi_{\rm
NP}(\tau,\mu)&&\mbox{(at QCD level)}\ .\end{align*}As the final
step, any ignorance about the contributions of higher states to
the spectral densities may be swept under the carpet by invoking
the assumption of (global) \emph{quark--hadron duality\/}
\cite{BSZ,MS,LM06}, postulating that above --- $\tau$-dependent
\cite{LMSET,LMSETa,LMSSET,LMSETb,LMSETc} --- effective thresholds
$s_{\rm eff}(\tau)$ the solely perturbative contributions to the
QCD-level correlator cancel the ones of all excited and continuum
states to the hadronic correlator:$$\int\limits_{s_{\rm
phys}}^\infty{\rm d}s\exp(-s\,\tau)\,\rho_{\rm cont}(s)\cong
\int\limits_{s_{\rm eff}(\tau)}^\infty{\rm d}s\exp(-s\,\tau)\,
\rho_{\rm pert}(s,\mu)\ .$$Allowing these cancellations to do
their job provides the QCD sum rule for the ground-state
hadron~$H$:$$f_H^2\exp(-M_H^2\,\tau)=\int\limits_{s_{\rm
th}}^{s_{\rm eff}(\tau)}{\rm d}s\exp(-s\,\tau)\,\rho_{\rm
pert}(s,\mu)+\Pi_{\rm NP}(\tau,\mu)\ .$$Eventually, after
determination of the $\tau$ dependence of the effective threshold
$s_{\rm eff}(\tau)$ along the lines of a well-developed
\cite{LMSAU,LMSAUa,LMSAUb,LMSAUc,MAU} algorithm, the decay
constant $f_H$ may be extracted from this QCD sum~rule~by
knowledge of the hadron's mass $M_H,$ or for the choice $\tau=0,$
that is, in the so-called local-duality~limit.

\section{Strong hadron decays: QCD sum rules from three-point
vertex functions}
\begin{figure}[ht]\sidecaption\centering
\includegraphics[scale=.23441]{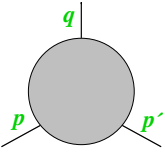}\caption{Three-point vertex
function $\Gamma(p,p',q)$ of three hadrons with momenta $p,$ $p',$
and $q.$}\label{3}\end{figure}

\emph{Mutatis mutandis\/}, hadron form factors
\cite{IS,NR} may be inferred from a three-point correlation
function$$\Gamma(p,p',q)\equiv\int{\rm d}^4x\,{\rm d}^4y\exp({\rm
i}\,p\,x-{\rm i}\,p'\,y)\,\langle\Omega|{\rm T}(j(x)\,J(0)\,j'(y))
|\Omega\rangle$$of three currents, $j(x),$ $j'(x),$ and $J(x),$
interpolating three hadrons, $H,$ $H',$ and $H_J,$ with decay
constants$$\langle\Omega|j(0)|H\rangle\propto f_H\ ,\qquad
\langle\Omega|j'(0)|H'\rangle\propto f_{H'}\ ,\qquad
\langle\Omega|J(0)|H_J\rangle\propto f_{H_J}\ ,$$masses
$M_H,M_{H'},M_{H_J},$ and momenta $p,p',q,$ respectively. Their
corresponding (simple) kinematics is illustrated by Fig.~\ref{3}.
For all external momenta timelike,
\emph{i.e.}, for $p^2>0,$ $p'^2>0,$ and $q^2>0,$~this~vertex
function develops a triple pole. One of these poles we attribute
to a form factor $F(q^2)$ of the hadron $H_J.$ The residue of the
latter pole at $q^2=M_{H_J}^2$ is defined by the strong coupling
$g_{HH'H_J}$ of the three~hadrons:$$\Gamma(p,p',q)=
\frac{f_H}{p^2-M_H^2}\,\frac{f_{H'}}{p'^2-M_{H'}^2}\,F(q^2)+\cdots\
,\qquad F(q^2)=\frac{f_{H_J}}{q^2-M_{H_J}^2}\,g_{HH'H_J}+\cdots\
.$$Again, complex analysis puts us in a position to construct a
representation of this three-point correlator $\Gamma(p,p',q)$ in
form of a double dispersion integral of the appropriate double
spectral density $\Delta(s,s',q^2)$:$$\Gamma(p,p',q)=
\int\frac{{\rm d}s}{s-p^2}\,\frac{{\rm d}s'}{s'-p'^2}\,
\Delta(s,s',q^2)\ .$$A double Borel transformation (from the
momentum variables $p^2$ and $p'^2$ to the Borel parameters $\tau$
and $\tau',$ respectively) \cite{IS,NR,BR} and the assumption that
quark--hadron duality holds in both the $H$ system~and the $H'$
system above continuum thresholds $s_{\rm eff}$ and $s'_{\rm
eff},$ respectively, yield the Borelized~QCD sum~rule
$$\exp(-M_H^2\,\tau)\exp(-M_{H'}^2\,\tau')\,F(q^2)=
\int\limits^{s_{\rm eff}}{\rm d}s\exp(-s\,\tau)
\int\limits^{s'_{\rm eff}}{\rm d}s'\exp(-s'\,\tau')\,
\Delta(s,s',q^2)\ .$$By focusing to the series expansion of the
perturbative part $\Gamma_{\rm pert}(p^2,p'^2,q^2)$ of the
correlator $\Gamma(p,p',q)$ $$\Gamma_{\rm
pert}(p^2,p'^2,q^2)=\Gamma_0(p^2,p'^2,q^2)+\alpha_{\rm
s}\,\Gamma_1(p^2,p'^2,q^2)+O(\alpha_{\rm s}^2)$$in powers of the
strong coupling $\alpha_{\rm s},$ let us now use this example to
try to work out the differences in the QCD sum-rule analysis of
ordinary hadrons, on the one hand, and of exotic hadrons, on
the~other~hand, as well as to cast some light on difficulties
encountered in such studies of strong decays of exotic states.

\section{Correlators of ordinary hadrons: strong
three-ordinary-meson couplings}Ordinary mesons, regarded as bound
states of a quark and an antiquark, can be interpolated by
suitable quark--antiquark currents $j(x)$ defined in terms of
quark fields $q_i(x)$ and generalized Dirac matrices~$\Gamma_A$:
$$j(x)\equiv\ :\!\!\bar q_i(x)\,\Gamma_A\,q_j(x)\!\!:\ ,\qquad
q_{i,j}=u,d,s,\dots\ ,\qquad A=1,2,\dots,16\ .$$

\begin{figure}[h]\sidecaption\centering\includegraphics
[scale=.269569]{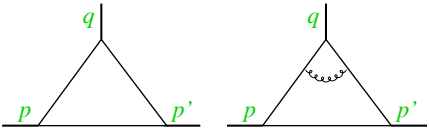}\caption{Perturbative expansion of
three-point correlation functions $\Gamma(p,p',q)$ of bilinear
quark currents interpolating ordinary mesons: leading-order
Feynman diagram (left) vs.\ one of the next-to-leading-order
Feynman diagrams (right).}\label{o}\end{figure}

\noindent Inspection of the perturbative expansion (Fig.~\ref{o})
shows that the leading-order spectral density and thus the form
factor $F(q^2)$ receive nonzero contributions from the one-loop
lowest-order Feynman diagram, which already provide the bulk of
the total three-point correlator \cite{LM12,LM11,BLM11a,BLM11b,
MSa,MSb}, whereas the contributions of higher-order Feynman
diagrams \cite{BBG,CFP,BOa,BOb} stay non-negligible, especially in
the large-$q^2$ range \cite{AMNa,AMNb,BLM08}.

\section{Correlators of one exotic and two ordinary hadrons:
exotic-meson decay}
\begin{figure}[h]\sidecaption\centering\includegraphics
[scale=.279563]{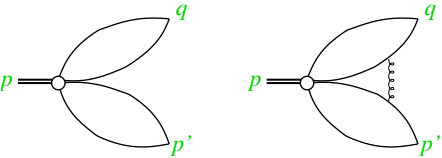}\caption{Perturbative expansion of
three-point correlation functions $\Gamma^{jj{\cal T}}_{\rm
QCD}(p^2,p'^2,q^2)$ of one exotic tetraquark current ${\cal T}$
and two ordinary bilinear quark currents $j$: lowest-order (left)
vs.\ one of the next-to-leading-order (right) diagrams.}\label{e}
\end{figure}

For states composed of four quarks, more
precisely, of two quarks and two antiquarks, the construction of
operators that interpolate such states is by no means unambiguous.
A given four-quark exotic~meson can be interpolated by, for
instance, tetraquark interpolating operators ${\cal T}(x)$
constructed from products of two quark--antiquark currents ${\cal
J}_{ij}(x),$ which, in turn, are defined by use of appropriate
combinations, generically labelled $\mathfrak{M},$ of Dirac and
colour matrices and covariant derivatives of the quark fields
$q_i(x)$:$${\cal T}(x)\equiv\ :\!\!{\cal J}_{12}(x)\,{\cal
J}_{34}(x)\!\!:\ ,\qquad{\cal J}_{ij}(x)\equiv\bar
q_i(x)\,\mathfrak{M}\,q_j(x)\ ,\qquad q_i=u,d,s,\dots\ .$$The
strong decays of any four-quark state, created from the QCD vacuum
$|\Omega\rangle$ by this operator, into two ordinary mesons can be
analysed by investigation of the three-point correlation function
$\Gamma^{jj{\cal T}}(x,y,z)$~of two currents $j(x)$ interpolating
the generated ordinary mesons and one of our tetraquark currents
${\cal T}(x)$:$$\Gamma^{jj{\cal T}}(x,y,z)\equiv
\langle\Omega|{\rm T}(j(x)\,j(y)\,{\cal T}(z))|\Omega\rangle\ .$$
The QCD-level evaluation of this correlation function reveals a
crucial difference in the QCD sum-rule description of the
characteristics of ordinary hadrons, on the one hand, and exotic
hadrons, on the other hand: By arguments based on diagrammatic
analysis, it is straightforward to convince oneself that, in a
perturbative expansion of the corresponding momentum-space
three-point correlator $\Gamma^{jj{\cal T}}_{\rm
QCD}(p^2,p'^2,q^2)$ in powers of the strong coupling $\alpha_{\rm
s}$ (cf.~Fig.~\ref{e}), the lowest-order contribution
$\Gamma^{jj{\cal T}}_0(p^2,p'^2,q^2),$ of order $O(\alpha_{\rm
s}^0),$ factorizes into the two-point correlation functions
$\Pi(p^2),$ \emph{i.e.}, exhibits the factorization property
$$\Gamma^{jj{\cal T}}_0(p^2,p'^2,q^2)=\Pi(p'^2)\,\Pi(q^2)\
.$$Consequently, the lowest-order contribution arises from
\emph{disconnected\/} diagrams.\footnote{This result is
corroborated by realizing that, upon Borel transformation with
respect to the exotic-hadron momentum $p^2,$ the leading-order,
disconnected contribution $\Gamma^{jj{\cal T}}_0(p^2,p'^2,q^2)$ to
the three-point correlator $\Gamma^{jj{\cal T}}_{\rm
QCD}(p^2,p'^2,q^2)$ gets mapped onto~zero.} Connected diagrams
start to contribute to the correlator $\Gamma^{jj{\cal T}}_{\rm
QCD}(p^2,p'^2,q^2)$ at perturbative order $O(\alpha_{\rm s})$ or
higher; this suggests to isolate the disconnected term from all
connected contributions $\Gamma^{jj{\cal T}}_{\rm
conn}(p^2,p'^2,q^2)$ to $\Gamma^{jj{\cal T}}_{\rm
QCD}(p^2,p'^2,q^2)$:$$\Gamma^{jj{\cal T}}_{\rm
QCD}(p^2,p'^2,q^2)=\Pi(p'^2)\,\Pi(q^2) +\alpha_{\rm
s}\,\Gamma^{jj{\cal T}}_{\rm conn}(p^2,p'^2,q^2)\ .$$

Unfortunately, however, \emph{disconnected\/} diagrams
bear no relationship at all to the bound states in the focus of
our interest; thus, they \emph{cannot\/} serve the purpose of
acquiring reliable information.\footnote{For related
considerations targeting the limit of the number of colours,
$N_{\rm c},$ becoming large, $N_{\rm c}\to\infty,$ consult
Refs.~\cite{SW,KP,CLa,CLb}.} Therefore, studies aiming at the
extraction of any features of exotic mesons, such as decay
amplitudes,~from QCD sum rules that rely exclusively on the
factorizable leading-order contribution to the relevant
correlation function must be doomed to fail. For some of all the
attempts in this latter direction, see
Refs.~\cite{NNL,NN,WH,D+,MN+}. Consequently, the investigation of
exotic states by QCD sum rules necessitates the consideration of
the contributions of \emph{connected\/} diagrams, arising, for
exotic states, at higher-than-trivial perturbative~order.

\section{Musings: prospects of elucidation of internal structure of
exotic mesons}Depending on the Lorentz nature of the employed
operators encoded in generalized Dirac matrices $\Gamma_A,$
$A=1,2,\dots,16,$ an ordinary meson $M$ may be interpolated by
only a comparatively small set of (linear combinations of)
bilinear quark currents $j_A(x)$ and therefore has only one or a
few decay~constants~$f_{M_A}$:$$\langle\Omega|j_A(0)|M\rangle
\propto f_{M_A}\ ,\qquad j_A(x)\equiv\ :\!\!\bar
q_1(x)\,\Gamma_A\,q_2(x)\!\!:\ ,\qquad q_{i,j}=u,d,s,\dots\ .$$In
contrast to this, an exotic hadron may be interpolated by a whole
plethora of interpolating operators: for instance, a tetraquark
meson $\Theta$ may be described by products of two
quark--antiquark currents $J(x),$ $$\langle\Omega|{\cal
M}(0)|\Theta\rangle\propto f_{\cal M}\ ,\qquad{\cal M}(x)\equiv\
:\!\!J_{12}(x)\,J_{34}(x)\!\!:\ ,\qquad J_{ij}(x)\equiv\bar
q_i(x)\,\Gamma_A\,q_j(x)\ ,$$or of a diquark current ${\cal
A}^a(x)$ and an antidiquark current $\bar{\cal A}^a(x),$ carrying
colour indices $a,b,c=1,2,3,$ $$\langle\Omega|{\cal
D}(0)|\Theta\rangle\propto f_{\cal D}\ ,\qquad{\cal D}(x)\equiv\
:\!\!{\cal A}^a_{12}(x)\,\bar{\cal A}^a_{34}(x)\!\!:\ ,\qquad{\cal
A}^a_{ij}(x)=\epsilon_{abc}\,q_{i,b}^{\rm
T}(x)\,\Gamma_A\,q_{j,c}(x)\ .$$Accordingly, a tetraquark meson
$\Theta$ may be characterized by a large number of decay constants
defined by the four-quark interpolating currents ${\cal
T}(x),{\cal M}(x),{\cal D}(x)$ and providing clues to its
internal~structure:\begin{itemize}\item On the one hand, if all
members $f_{\cal M}$ of the set of decay constants caused by
quark--antiquark currents prove to be of low numerical relevance,
$f_{\cal M}\approx0\ \forall{\cal M},$ we may be tempted to call
the exotic meson $\Theta$ a colour-neutral, of course, bound state
of a colour-triplet diquark and a colour-antitriplet
antidiquark.\item On the other hand, if all elements $f_{\cal D}$
of the set of decay constants of (anti-)diquark-current origin~are
comparatively tiny, $f_{\cal D}\approx0\ \forall{\cal D},$ we will
tend to regard the exotic meson $\Theta$ as a molecular-type bound
state built up by two colourless mesons created from the vacuum by
their interpolating currents~$J(x).$\end{itemize}Needless to say,
any operator with the appropriate or desired quantum numbers may
be adopted for the interpolation of, and can therefore potentially
contribute to, the hadronic state under consideration. An exotic
meson, for example, might be found to consist of both a two-quark
and a four-quark component. In addition, Fierz transformations
offer the possibility to reshuffle the order of the quark field
operators in linear combinations of four-quark interpolating
currents and to switch between diquark--antidiquark and
meson--meson structures \cite{jaffe}. All this contributes to
blurring the emerging picture of exotic states.

Theoretically, decay constants of the kind introduced above may be
approached by QCD sum rules based on the two-point correlation
functions of suitable tetraquark interpolating operators
$\theta(x)$~\cite{FNR,A+}$$\left\langle\Omega\left|{\rm T}\!\left
(\theta(x)\,\theta^\dag(0)\right)\right|\Omega\right\rangle\,,\qquad
\theta(x)\in\{{\cal T}(x),{\cal M}(x),{\cal D}(x),\dots\}\ .$$In
this context, we face, however, the problem that, similar to the
case of three-point correlators, not all contributions to these
two-point correlation functions are related to the properties of
the corresponding exotic states. Notwithstanding this, we are
confident \cite{LM@} that the QCD sum-rule formalism enables~us to
derive exotic-hadron decay constants in a similar manner as in the
case of ordinary hadrons \cite{LMSDC,LMSDCa,LMSDCb,LMSDCc,LMSRc}.

\acknowledgement{D.~M.~would like to express gratitude for support
by the Austrian Science Fund (FWF) under project P29028-N27.}\vfil

\end{document}